\documentclass[aps,preprint,amsmath,amssymb]{revtex4}
\usepackage{graphicx}
\usepackage{dcolumn}
\usepackage{bm}
\usepackage{amssymb}
\usepackage{amsmath}

\begin{document}


\title{Statistical mechanics characterization of neuronal mosaics}

\author{Luciano da Fontoura Costa} 
\email{luciano@if.sc.usp.br}
 \affiliation{Instituto de F\'{\i}sica de S\~{a}o Carlos, Universidade
 de S\~{a}o Paulo, Av. Trabalhador S\~{a}o Carlense 400, Caixa Postal
 369, CEP 13560-970, S\~{a}o Carlos, S\~ao Paulo, Brazil}

\author{Fernando Rocha and Silene Maria Ara\'ujo de Lima}
\affiliation{Centro de Ci\^encias Biol\'ogicas, Departamento de
Fisiologia, Universidade Federal do Par\'a, Campus Universit\'ario do
Guam\'a, Rua Augusto Corr\^ea 01, CEP 66075-000, Bel\'em, Par\'a,
Brazil}

\date{3rd November 2004}

\begin{abstract}
The spatial distribution of neuronal cells is an important requirement for
achieving proper neuronal function in several parts of the nervous
system of most animals.  For instance, specific distribution of
photoreceptors and related neuronal cells, particularly the ganglion
cells, in mammal's retina is required in order to properly sample the
projected scene.  This work presents how two concepts from the areas
of statistical mechanics and complex systems, namely the
\emph{lacunarity} and the \emph{multiscale entropy} (i.e. the
entropy calculated over progressively diffused representations of the
cell mosaic), have allowed effective characterization of the spatial
distribution of retinal cells.
\end{abstract}


\maketitle

Although a great part of neuronal function is a consequence of the
distribution and adaptation of the involved synaptic weights, the
topographical nature of sensory spaces found in several animals
implies the spatial distribution of neuronal cells to become an
important element for achieving proper overall behavior of neuronal
subsystems \cite{Jacobs:1993,Szel_etal:2000}.  The topographical
organization of neuronal systems and mappings is characterized by the
preservation of local adjacencies of the represented and mapped points
(e.g. \cite{Costa_Diambra:2004}). In the retina, for instance, the
photoreceptors and associated neuronal cells are spatially organized
as \emph{mosaics} so as to obtain proper sampling of the projected
images, which may depend on the specific spectral range and type of
visual operation (e.g. central/peripheric and chromatic/scotopic).
More specifically, while such a distribution tends to lead to locally
uniform distribution of receptive fields, with just the right amount
of overlap, the overall distribution varies with the eccentricity,
becoming less dense at the paraphoveal regions and periphery of the
retina.

At the same time, retinae of several species involve more than one
type of photoreceptors, specialized at some particular wavelength
interval, implying additional restrictions to the spatial
distributions in each respective mosaic.  Combined with the close
relationship between neuronal structure and function, such specific
demands makes the problem of mosaic spatial characterization
particularly interesting from both the mathematical and physiological
points of view.  At the same time, most related researches reported in
the specific literature are almost invariably related to the
application of nearest-neighbor distances between the mosaic
elements.  Although such a measurement is interesting and has paved
the way to several advances, the consideration of additional,
complementary features to describe the spatial organization has the
potential to provide more information about the investigated systems.

This article describes the application, to our best knowledge for the
first time, of two concepts derived from statistical mechanics and
complex systems --- namely the lacunarity and multiscale entropy ---
to the characterization of retinal mosaics.  

Introduced \cite{Gefen_etal:1983, Hovi_etal:1996, Allain_Cloitre:1991,
Einstein:1998} in order to complement the degenerated characterization
provided by the fractal dimension, the \emph{lacunarity} provides a
measurement of the \emph{translational invariance} of the analyzed
spatial distributions.  Therefore, the higher the value of the
lacunarity, the less translationally invariant is the system.  Let the
sum of the mosaic values under the circular area of radius $r$
centered at each point $(x,y)$, represented as $\phi$, be

\begin{equation}
  s_r(x,y)= \sum_{\phi} a(x,y)
\end{equation}

where $a$ is the matrix representing the mosaic.  One of the
frequently adopted definitions of lacunarity is as follows

\begin{equation}
  L(r) = \frac{\sigma_r^2}{\mu_r^2}
\end{equation}

where $\mu_r$ and $\sigma_r$ are the mean and standard deviation of
$s_r(x,y)$ taken along the $xy$ domain.  Observe that the lacunarity
maps the image into the vector $L(r)$, $r = 1, 2, \ldots, r_{max}$.
For enhanced computation sake, the lacunarity was obtained by
convolving, in the spectral domain, the original mosaic with disks of
increasing radius.

The multiscale entropy \cite{entr_mult:1998, CostaCesar:2001} can be
used to complement the degeneracy of the traditional entropy, which is
invariant to any spatial permutation of the mosaic values (the entropy
ultimately depends only on the image value histogram which, as a first
order statistics, does not depend on the spatial positions).  More
specifically, the multiscale entropy can be defined as follows

\begin{equation}
  E_{\sigma} = - \sum_{v} p_{\sigma}(v) log(p_{\sigma}(v))
\end{equation}

where $p_{\sigma}(v)$ is the histogram of the mosaic values after
diffusion with standard deviation $\sigma$ (e.g. Gaussian smoothing,
which is equivalent to linear isotropic diffusion).  The diffused
versions of the original mosaic images were obtained through the
convolution, in the spectral domain, of the original mosaic image with
normal density functions for increasing values of $\sigma$
\cite{CostaCesar:2001}.

\begin{figure*}
\begin{center}
\includegraphics[scale=0.4]{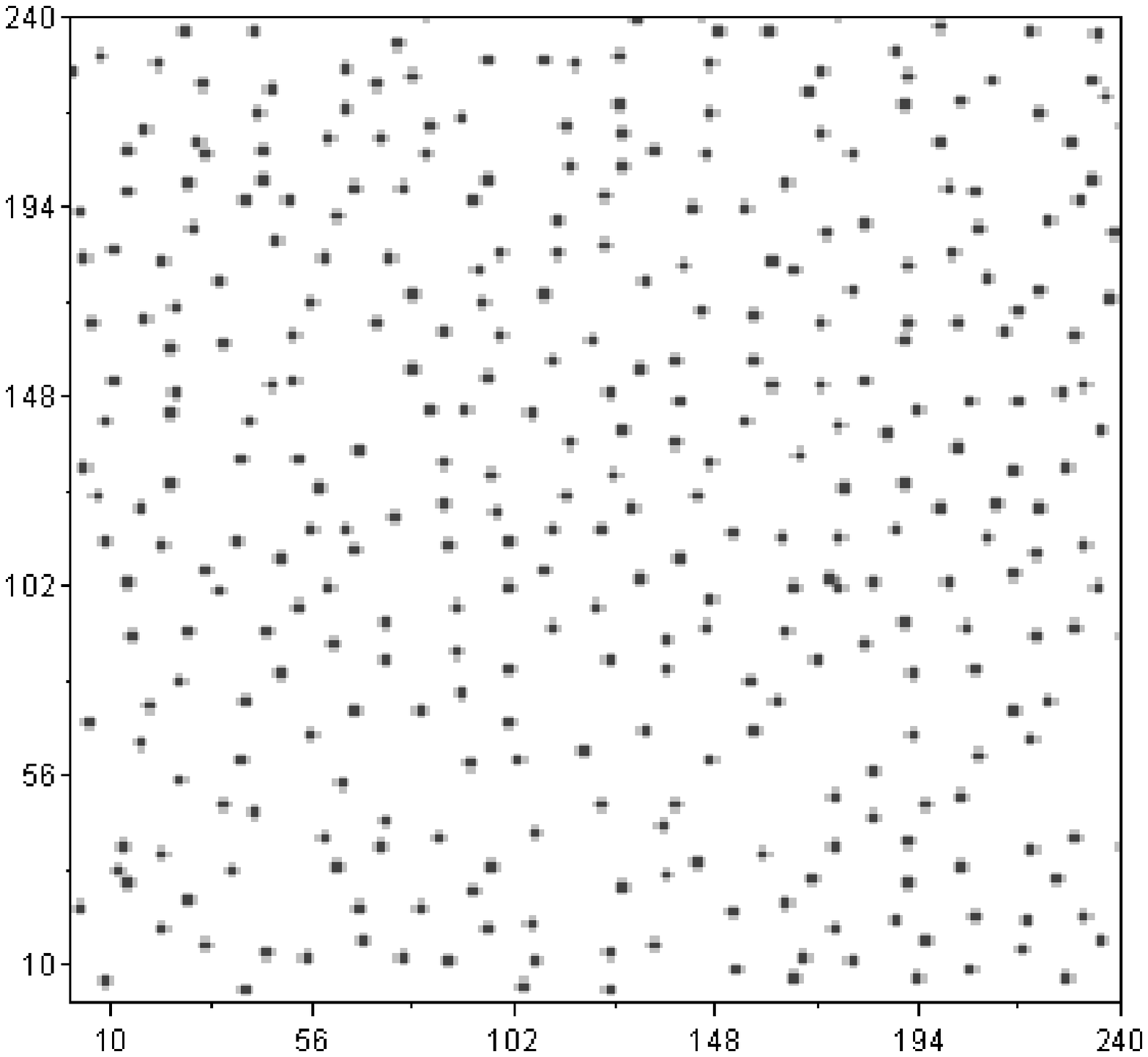} 
\includegraphics[scale=0.4]{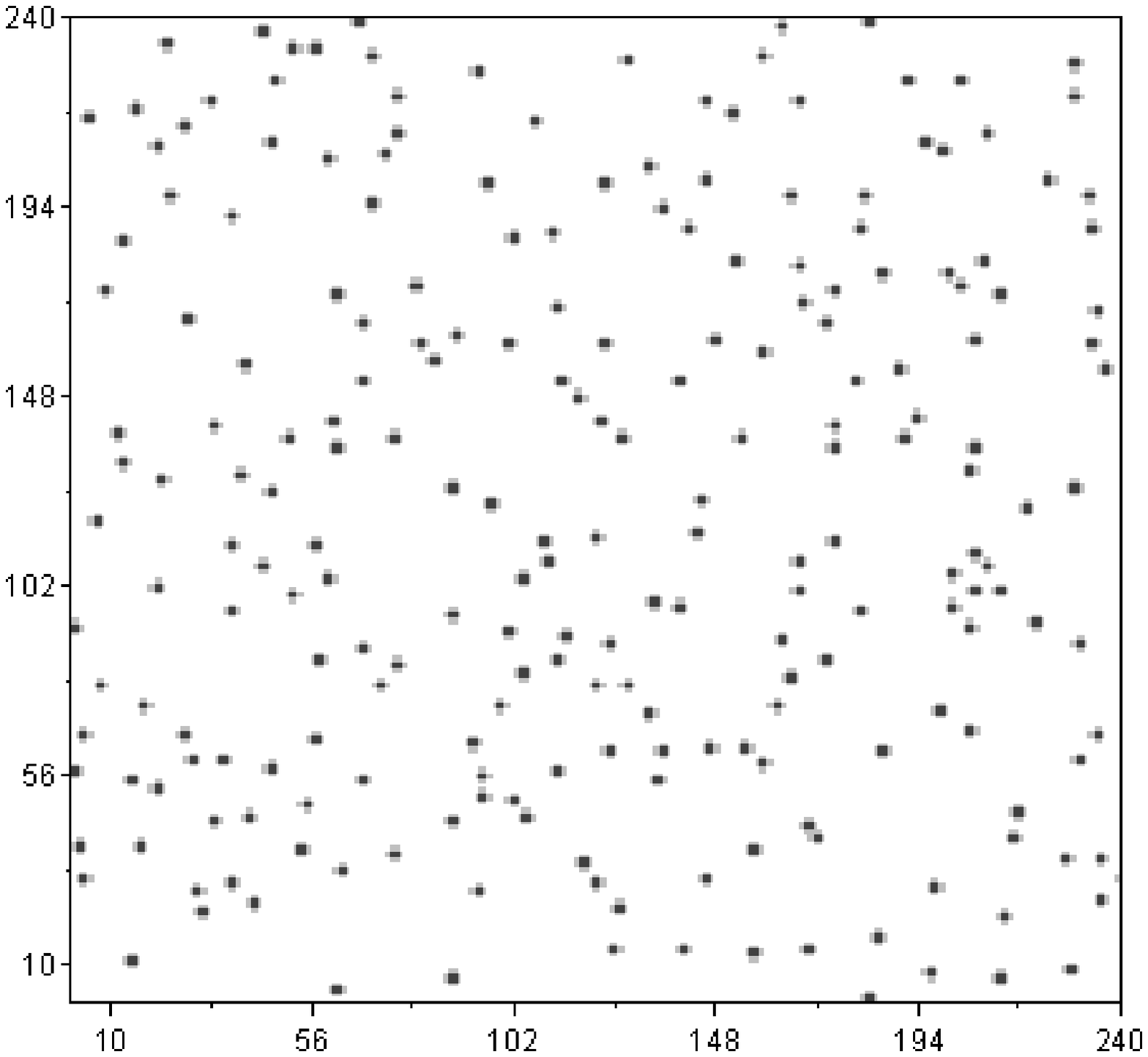} \\
(a)  \hspace{7cm} (b) \\
\includegraphics[scale=0.7]{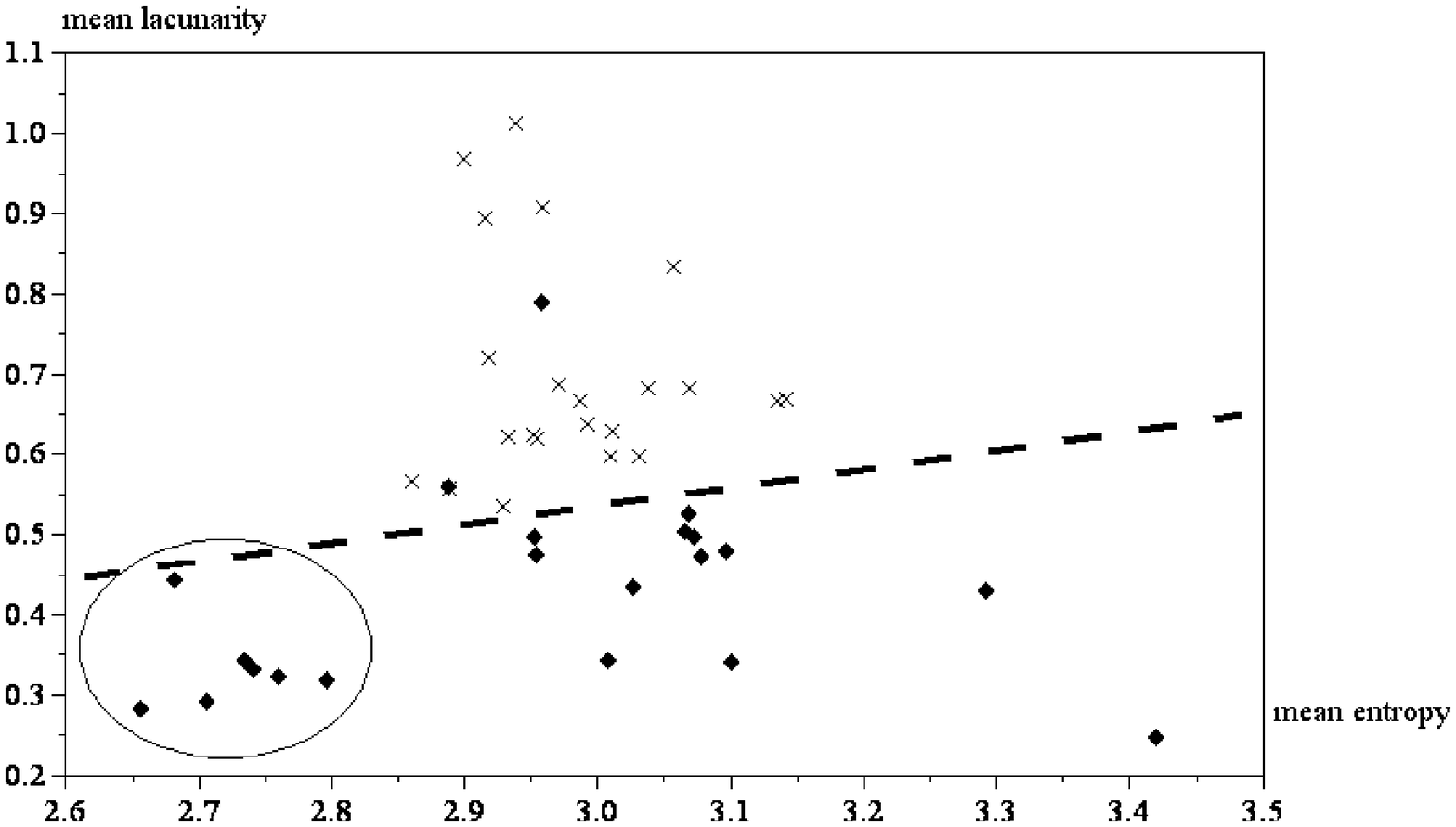} \\
(c)
\end{center}
\caption{\label{scatter} Figure 1.}
\end{figure*}

The remainder of this article presents the application of the two
above measurements to the characterization of Agouti (\emph{Dasiprocta
agouti}) retinal mosaics, which involve two types of cells (M/L and S)
characterized by different pigments.  The process of enucleation and
fixation of the retinal material followed the protocols described
elsewhere \cite{Ahnelt_etal:1995}. Isolated retinas were prepared as
whole mounts and were then labeled with different antibodies (JH-455,
against S cone pigments and JH-492 against M/L cone pigments; kindly
provided by Dr. J. Nathans) largely according to the procedures
described elsewhere (*). The whole-mounted retinas were examined and
scanned using differential interference contrast (DIC) with the aid of
an optical microscope (LEICA DM) equipped with a high-resolution video
camera (AxionCam MRm). Starting from the optic nerve, images of S and
M/L cones were acquired along the vertical-dorsal axis. Using 40X oil
immersion objective, 25 fields taken with $250 \times 250\mu m$ for M/L
cones and 23 fields with $500 \times 500 \mu m$ for S cones were
digitalized. The raw images of the cones were captured at the level of
the inner segments. For analysis, $x$ and $y$ coordinates were
identified with Scion Image Software (ScionCorp).  Digital binary
images were then obtained (cells are represented by 1 and absence of
cells by 0) and represented as matrices of fixed size $250 \time 250$
(the coordinates of S cones were divided by 2) in order to normalize
finite-size effects.  Typical examples of the M/L and S cell
distributions are provided in Figure~\ref{scatter}(a) and (b),
respectively.

Figure~\ref{scatter}(c) presents the scatterplot considering the mean
lacunarity and mean multiscale-entropy obtained for each mosaic.  It
is clear from this figure that both M/L and S cells resulted linearly
separated, in the sense that most cases above the dashed line are S
cells.  The two exceptions were found to have uncommon features.  All
cells below the line are M/L cells.  Because the separation line is
almost horizontal and orthogonal to the $y$-axis, it follows that such
a discrimination is obtained mostly as a consequence of the mean
lacunarity measurements.  Indeed, as can be inferred from
Figures~\ref{scatter}(a) and (b), the M/L mosaics tend to be spatially
more uniform, implying higher translational invariance.  On the other
hand, the S mosaics tend to have greater variation as for the size and
shape of their void regions, leading to higher lacunarity.  Although
the mean multiscale entropy did not contribute to separating between M/L
and S mosaics, it did produce an isolated cluster at the left-hand
side of the figure, which is marked by the ellipse.  The mosaics in
that cluster have been verified to have a smaller number of cells, a
property which was properly reflected by the multiscale entropy.

All in all, we have shown that lacunarity and multiscale entropy
measurements derived from digital images of retinal mosaics can
provide valuable information about the spatial distribution of the
involved cells.  While the mean lacunarity accounted for an almost
perfect separation of the two types of mosaics, the mean multiscale
entropy allowed the identification of a separated cluster of M/L
mosaics.  Such results provide circumstantial evidence in favor of the
fact that irregular rather than regular patterns appear to be the rule
among mammals with S cone mosaics~\cite{Ahnelt_Kolb:2000}.  Random
distribution in the S cone mosaics have been identified in species
like marsupials, rabbits, cats, horses, rats, cheetahs and guinea
pigs~\cite{Szel_etal:1992, Sandmann_etal:1996, Ahnelt_etal:2000,
Ahnelt_etal:1995, Szel_Rohlich:1992} which, like the agouti, have
rod-dominant retina.  On the other hand, regular S mosaics are found
only in animals with cone-dominant retina~\cite{Muller_Peichl:1989,
Ahnelt:1985, Long_Fisher:1983}, possibly accounting for enhanced visual
sampling properties compatible with diurnal arboreal habitats.

\begin{acknowledgments}

Luciano da F. Costa thanks HFSP RGP39/2002, FAPESP (proc. 99/12765-2)
and CNPq (proc. 3082231/03-1) for financial support.

\end{acknowledgments}

\pagebreak

.

\pagebreak 

List of Captions:

Typical configurations of M/L (a) and S (b)
mosaics, and the scatterplot obtained for mean lacunarity and mean
entropy (c).  M/L and S mosaics are represented by crosses and diamonds,
respectively.

\pagebreak

\bibliography{mosaics}

\begin{thebibliography}{18}
\expandafter\ifx\csname natexlab\endcsname\relax\def\natexlab#1{#1}\fi
\expandafter\ifx\csname bibnamefont\endcsname\relax
  \def\bibnamefont#1{#1}\fi
\expandafter\ifx\csname bibfnamefont\endcsname\relax
  \def\bibfnamefont#1{#1}\fi
\expandafter\ifx\csname citenamefont\endcsname\relax
  \def\citenamefont#1{#1}\fi
\expandafter\ifx\csname url\endcsname\relax
  \def\url#1{\texttt{#1}}\fi
\expandafter\ifx\csname urlprefix\endcsname\relax\def\urlprefix{URL }\fi
\providecommand{\bibinfo}[2]{#2}
\providecommand{\eprint}[2][]{\url{#2}}

\bibitem[{\citenamefont{Jacobs}(1993)}]{Jacobs:1993}
\bibinfo{author}{\bibfnamefont{G.~H.} \bibnamefont{Jacobs}},
  \bibinfo{journal}{Biol. Rev. Camb. Philos. Soc.}
  \textbf{\bibinfo{volume}{68}}, \bibinfo{pages}{413} (\bibinfo{year}{1993}).

\bibitem[{\citenamefont{Szel et~al.}(2000)\citenamefont{Szel, Lukats, Fetke,
  Szepessy, and Rohlich}}]{Szel_etal:2000}
\bibinfo{author}{\bibfnamefont{A.}~\bibnamefont{Szel}},
  \bibinfo{author}{\bibfnamefont{A.}~\bibnamefont{Lukats}},
  \bibinfo{author}{\bibfnamefont{T.}~\bibnamefont{Fetke}},
  \bibinfo{author}{\bibfnamefont{Z.}~\bibnamefont{Szepessy}}, \bibnamefont{and}
  \bibinfo{author}{\bibfnamefont{P.}~\bibnamefont{Rohlich}},
  \bibinfo{journal}{J. Opt. Soc. Ame. A Opt. Image Sci. Vis.}
  \textbf{\bibinfo{volume}{17}}, \bibinfo{pages}{568} (\bibinfo{year}{2000}).

\bibitem[{\citenamefont{da~F.~Costa and Diambra}(2004)}]{Costa_Diambra:2004}
\bibinfo{author}{\bibfnamefont{L.}~\bibnamefont{da~F.~Costa}} \bibnamefont{and}
  \bibinfo{author}{\bibfnamefont{L.}~\bibnamefont{Diambra}},
  \bibinfo{journal}{Phys. Rev. E}  (\bibinfo{year}{2004}),
  \bibinfo{note}{cond-mat/0305010, accepted}.

\bibitem[{\citenamefont{Gefen et~al.}(1983)\citenamefont{Gefen, Meir,
  Mandelbrot, and Aharony}}]{Gefen_etal:1983}
\bibinfo{author}{\bibfnamefont{Y.}~\bibnamefont{Gefen}},
  \bibinfo{author}{\bibfnamefont{Y.}~\bibnamefont{Meir}},
  \bibinfo{author}{\bibfnamefont{B.~B.} \bibnamefont{Mandelbrot}},
  \bibnamefont{and} \bibinfo{author}{\bibfnamefont{A.}~\bibnamefont{Aharony}},
  \bibinfo{journal}{Phys. Rev. Letts.} \textbf{\bibinfo{volume}{50}},
  \bibinfo{pages}{145} (\bibinfo{year}{1983}).

\bibitem[{\citenamefont{Hovi et~al.}(1996)\citenamefont{Hovi, Aharony,
  Stauffer, and Mandelbrot}}]{Hovi_etal:1996}
\bibinfo{author}{\bibfnamefont{J.-P.} \bibnamefont{Hovi}},
  \bibinfo{author}{\bibfnamefont{A.}~\bibnamefont{Aharony}},
  \bibinfo{author}{\bibfnamefont{D.}~\bibnamefont{Stauffer}}, \bibnamefont{and}
  \bibinfo{author}{\bibfnamefont{B.~B.} \bibnamefont{Mandelbrot}},
  \bibinfo{journal}{Phys. Rev. Letts.} \textbf{\bibinfo{volume}{77}},
  \bibinfo{pages}{877} (\bibinfo{year}{1996}).

\bibitem[{\citenamefont{Allain and Cloitre}(1991)}]{Allain_Cloitre:1991}
\bibinfo{author}{\bibfnamefont{C.}~\bibnamefont{Allain}} \bibnamefont{and}
  \bibinfo{author}{\bibfnamefont{M.}~\bibnamefont{Cloitre}},
  \bibinfo{journal}{Phys. Rev. A} \textbf{\bibinfo{volume}{44}},
  \bibinfo{pages}{3552} (\bibinfo{year}{1991}).

\bibitem[{\citenamefont{Einstein et~al.}(1998)\citenamefont{Einstein, Wu, and
  Gil}}]{Einstein:1998}
\bibinfo{author}{\bibfnamefont{A.~J.} \bibnamefont{Einstein}},
  \bibinfo{author}{\bibfnamefont{H.~S.} \bibnamefont{Wu}}, \bibnamefont{and}
  \bibinfo{author}{\bibfnamefont{J.}~\bibnamefont{Gil}},
  \bibinfo{journal}{Phys. Rev. Letts.} \textbf{\bibinfo{volume}{80}},
  \bibinfo{pages}{387} (\bibinfo{year}{1998}).

\bibitem[{\citenamefont{da~F.~Costa and Jr}(2001)}]{CostaCesar:2001}
\bibinfo{author}{\bibfnamefont{L.}~\bibnamefont{da~F.~Costa}} \bibnamefont{and}
  \bibinfo{author}{\bibfnamefont{R.~M.~C.} \bibnamefont{Jr}},
  \emph{\bibinfo{title}{Shape Analysis and Classification: Theory and
  Practice}} (\bibinfo{publisher}{CRC Press}, \bibinfo{address}{Boca Raton},
  \bibinfo{year}{2001}).

\bibitem[{\citenamefont{Bruno et~al.}(1998)\citenamefont{Bruno, Cesar,
  Consularo, and da~F.~Costa}}]{entr_mult:1998}
\bibinfo{author}{\bibfnamefont{O.~M.} \bibnamefont{Bruno}},
  \bibinfo{author}{\bibfnamefont{R.~M.} \bibnamefont{Cesar}},
  \bibinfo{author}{\bibfnamefont{L.~A.} \bibnamefont{Consularo}},
  \bibnamefont{and}
  \bibinfo{author}{\bibfnamefont{L.}~\bibnamefont{da~F.~Costa}}, in
  \emph{\bibinfo{booktitle}{Proc. International Symposium on Computer Graphics,
  Image Processing and Vision, SIBGRAPI}} (\bibinfo{publisher}{IEEE Computer
  Society Press}, \bibinfo{address}{Rio de Janeiro, RJ}, \bibinfo{year}{1998}),
  vol.~\bibinfo{volume}{18}, pp. \bibinfo{pages}{363--370}.

\bibitem[{\citenamefont{Ahnelt et~al.}(1995)\citenamefont{Ahnelt, Hokoc, and
  Rohlich}}]{Ahnelt_etal:1995}
\bibinfo{author}{\bibfnamefont{P.~K.} \bibnamefont{Ahnelt}},
  \bibinfo{author}{\bibfnamefont{J.~N.} \bibnamefont{Hokoc}}, \bibnamefont{and}
  \bibinfo{author}{\bibfnamefont{P.}~\bibnamefont{Rohlich}},
  \bibinfo{journal}{Vis. Neurosc.} \textbf{\bibinfo{volume}{12}},
  \bibinfo{pages}{793} (\bibinfo{year}{1995}).

\bibitem[{\citenamefont{Ahnelt and Kolb}(2000)}]{Ahnelt_Kolb:2000}
\bibinfo{author}{\bibfnamefont{P.~K.} \bibnamefont{Ahnelt}} \bibnamefont{and}
  \bibinfo{author}{\bibfnamefont{H.}~\bibnamefont{Kolb}},
  \bibinfo{journal}{Prog. Retin. Eye Res.} \textbf{\bibinfo{volume}{19}},
  \bibinfo{pages}{711} (\bibinfo{year}{2000}).

\bibitem[{\citenamefont{Sandmann et~al.}(1996)\citenamefont{Sandmann, Boycott,
  and Peichl}}]{Sandmann_etal:1996}
\bibinfo{author}{\bibfnamefont{D.}~\bibnamefont{Sandmann}},
  \bibinfo{author}{\bibfnamefont{B.~B.} \bibnamefont{Boycott}},
  \bibnamefont{and} \bibinfo{author}{\bibfnamefont{L.}~\bibnamefont{Peichl}},
  \bibinfo{journal}{J. Neurosc.} \textbf{\bibinfo{volume}{16}},
  \bibinfo{pages}{3381} (\bibinfo{year}{1996}).

\bibitem[{\citenamefont{Szel et~al.}(1992)\citenamefont{Szel, Rohlich, Caffe,
  Juliusson, Aguirre, and van Veen}}]{Szel_etal:1992}
\bibinfo{author}{\bibfnamefont{A.}~\bibnamefont{Szel}},
  \bibinfo{author}{\bibfnamefont{P.}~\bibnamefont{Rohlich}},
  \bibinfo{author}{\bibfnamefont{A.~R.} \bibnamefont{Caffe}},
  \bibinfo{author}{\bibfnamefont{B.}~\bibnamefont{Juliusson}},
  \bibinfo{author}{\bibfnamefont{G.}~\bibnamefont{Aguirre}}, \bibnamefont{and}
  \bibinfo{author}{\bibfnamefont{T.}~\bibnamefont{van Veen}},
  \bibinfo{journal}{J. Comp. Neurol.} \textbf{\bibinfo{volume}{325}},
  \bibinfo{pages}{327} (\bibinfo{year}{1992}).

\bibitem[{\citenamefont{Ahnelt et~al.}(2000)\citenamefont{Ahnelt, Fernandez,
  Martinez, and Kubber-Heiss}}]{Ahnelt_etal:2000}
\bibinfo{author}{\bibfnamefont{P.~K.} \bibnamefont{Ahnelt}},
  \bibinfo{author}{\bibfnamefont{E.}~\bibnamefont{Fernandez}},
  \bibinfo{author}{\bibfnamefont{O.}~\bibnamefont{Martinez}}, \bibnamefont{and}
  \bibinfo{author}{\bibfnamefont{A.}~\bibnamefont{Kubber-Heiss}},
  \bibinfo{journal}{J. Opt. Soc. Am. A} \textbf{\bibinfo{volume}{17}},
  \bibinfo{pages}{580} (\bibinfo{year}{2000}).

\bibitem[{\citenamefont{Szel and Rohlich}(1992)}]{Szel_Rohlich:1992}
\bibinfo{author}{\bibfnamefont{A.}~\bibnamefont{Szel}} \bibnamefont{and}
  \bibinfo{author}{\bibfnamefont{P.}~\bibnamefont{Rohlich}},
  \bibinfo{journal}{Exp. Eye Res.} \textbf{\bibinfo{volume}{55}},
  \bibinfo{pages}{47} (\bibinfo{year}{1992}).

\bibitem[{\citenamefont{Muller and Peichl}(1989)}]{Muller_Peichl:1989}
\bibinfo{author}{\bibfnamefont{B.}~\bibnamefont{Muller}} \bibnamefont{and}
  \bibinfo{author}{\bibfnamefont{L.}~\bibnamefont{Peichl}},
  \bibinfo{journal}{J. Comp. Neurol.} \textbf{\bibinfo{volume}{282}},
  \bibinfo{pages}{581} (\bibinfo{year}{1989}).

\bibitem[{\citenamefont{Ahnelt}(1985)}]{Ahnelt:1985}
\bibinfo{author}{\bibfnamefont{P.~K.} \bibnamefont{Ahnelt}},
  \bibinfo{journal}{Vis. Res.} \textbf{\bibinfo{volume}{25}},
  \bibinfo{pages}{1557} (\bibinfo{year}{1985}).

\bibitem[{\citenamefont{Long and Fisher}(1983)}]{Long_Fisher:1983}
\bibinfo{author}{\bibfnamefont{K.~O.} \bibnamefont{Long}} \bibnamefont{and}
  \bibinfo{author}{\bibfnamefont{S.~K.} \bibnamefont{Fisher}},
  \bibinfo{journal}{J. Comp. Neurol.} \textbf{\bibinfo{volume}{221}},
  \bibinfo{pages}{329} (\bibinfo{year}{1983}).

\end{thebibliography}

\end{document}